\documentclass[%
 reprint,
 amsmath,amssymb,
 aps,
]{revtex4-1}

\usepackage{graphicx}
\usepackage{dcolumn}
\usepackage{bm}
\usepackage{hyperref}
\usepackage{xcolor}

\usepackage{amsfonts}
\usepackage{xcolor}

\newcommand{\ket}[1]{{\left\vert{#1}\right\rangle}}
\newcommand{\braket}[2]{{\langle {#1}\!\mid\!{#2} \rangle}}

\newtheorem{definition}{Definition}

\newtheorem{property}{Property}
\newtheorem{example}{Example}

\begin{document}

\preprint{APS/123-QED}

\title{Quantum hashing via single-photon states with orbital angular momentum}

\author{D.A. Turaykhanov$^{1}$}
\email{7intur@gmail.com}
\author{D.O. Akat'ev$^{1}$, A.V. Vasiliev$^{1,2}$, F.M. Ablayev$^{1,2}$}
\email{fablayev@gmail.com}
\author{A.A. Kalachev$^{1}$}
\email{a.a.kalachev@mail.ru}
\address{$^{1}$ Zavoisky Physical-Technical Institute, Kazan Scientific Center of the Russian Academy of Sciences, 10/7 Sibirsky Tract, Kazan 420029, Russia}%

\address{$^{2}$ Kazan Federal University, 18 Kremlyovskaya street, Kazan 420008, Russia}%

\date{\today}

\begin{abstract}
Quantum hashing is a promising generalization of the cryptographic hashing concept on the quantum domain. In this paper, we construct a quantum hash via a sequence of single-photon states and perform a proof-of-principle experiment using orbital angular
momentum (OAM) encoding.
We experimentally verify the collision resistance of the quantum hash function depending on the number of qubits in use. Based on these results, we conclude that theoretical estimates are confirmed for different bases of OAM states and the proposed technique can be useful in computational and cryptographic scenarios. The possibility of multiplexing different OAM bases can make this approach even more efficient.
\end{abstract}

\maketitle


\section{\label{sec:level1}Introduction}

Hashing is an important tool in theoretical and applied computer science. In general, the main task of hashing is to reliably present information (stings over some finite alphabet) in compressed form (hash) that allows one to distinguish the initial stings effectively. Such a mapping is known as a hash function and is widely used in various data storage and retrieval applications to reduce the data access time. On top of that, hashing is widely used in many parts of modern cryptography, such as verification of message integrity, digital signatures, fingerprinting, etc. \cite{Paar-Pelzl:2009:Understanding-Cryptography,Katz-Lindell:2014:Modern-Cryptography}. For cryptographic applications, hash functions should satisfy two main properties: one-way property and collision resistance. The first means that to find a hash value for an arbitrary input string should be a simple problem, but to restore an input from its hash should be a computationally hard one. The second property means that the situation when two different inputs have the same hash (such a situation is called a collision) is infeasible. In other words, to find a collision for cryptographic hash functions should also be a computationally hard problem. Note that collisions for hash functions mostly occur just because long inputs are compressed into short outputs.

A promising generalization of the cryptographic hashing concept in the quantum domain -- a quantum hashing -- has been suggested in \cite{Ablayev-Vasiliev:2014:Crypto-Q-Hashing}. In this case, the hash function is defined as a mapping of classical input strings into quantum states. Such a mapping should satisfy the following two main properties: (i) long strings $w$ should be mapped to quantum states $\ket{\psi(w)}$ composed of a small number of qubits; (ii) for arbitrary two different inputs $w,w'$ the fidelity between their quantum hashes $\ket{\psi(w)}, \ket{\psi(w')}$ should be small. The first property provides a good one-way function -- due to the fundamental Holevo theorem one cannot extract more information from a quantum state than the number of its qubits. The second property provides good collision resistance -- probability to find a collision is bounded by the value of fidelity. Such functions can be used as a basic ingredient for various quantum cryptography protocols: from privacy amplification in quantum key distribution to quantum fingerprinting, quantum signatures, etc. In papers \cite{Ablayev-Vasiliev:2014:Crypto-Q-Hashing,Ablayev-et-al:2016:Quantum-Fingerprinting-Hashing}, several mathematical constructions  of quantum hash functions that satisfy properties mentioned above were presented, in particular the combinatorial construction known as ''small-biased set''.  In addition, the  quantum fingerprinting, which also compresses classical inputs to quantum states,  has also been defined \cite{Buhrman:2001:Fingerprinting}. The proposed constructions of quantum hashing and quantum fingerprinting are maximally efficient if completely entangled quantum states are involved, which are hard to prepare. Therefore, it is interesting to study other variants of quantum hashing that minimising quantum state engineering.

In the present paper, we develop a new construction of a quantum hash as a sequence of single-photon states and implement a proof-of-principle experiment using orbital angular momentum encoding.
We map an input string $w$ (we also consider $w$ as a number for the technical convenience) on a quantum state $\ket{\psi(w)}$ that consists of a sequence of single isolated qubits. The main idea of quantum hashing here is that the state of each qubit is determined (in a specific way) by a whole input  $w$. The advantage of such an approach is that destruction of some of the qubits has no negative effect on the information about $w$. In doing so, we use single-photon qubits encoded via orbital angular momentum (OAM) that form a promising platform for optical quantum technologies \cite{Flamini:2019bp}. In particular, they have been used in a series of experiments on free-space quantum communication \cite{DAmbrosio:2012tr,Vallone2014,krenn2014communication,krenn2015twisted,Mirhosseini:2015fy,Goyal:2016fw,Sit:2017ie,Mirhosseini:2015fy,Bouchard2018experimental}. In the context of quantum hashing, two points are especially worth noting: the possibility of high-speed data transmission by means of OAM-based multiplexing \cite{wang2012terabit,huang2014,Lei:2015kg} and the possibility of preparing high-dimensional quantum systems using OAM encoding \cite{Mirhosseini:2015fy,Sit:2017ie,Bouchard2018experimental,Erhard:2018hp}. The former allows multiplexing of quantum hashing while the latter can be used for increasing its efficiency without entanglement. Here we implement quantum hashing using a sequence of single-photon qubits with different values of OAM, which may be considered as a necessary and vital step towards realizing the first possibility.

\section{Quantum Hashing}

We start with recalling the definition of a quantum hash function and its properties.

We define a classical-quantum (or just quantum) function $\psi$ over the finite set  of classical words (i.e. finite sequences of input symbols) $\mathbb X$ as
\begin{equation}\label{qf}
 \psi :  \mathbb X \to ({\cal H}^2)^{\otimes s}.
 \end{equation}

In other words a quantum function $\psi$ encodes an input $w\in \mathbb X$ into an $s$-qubit quantum state $\ket{\psi(w)}$. In order to construct cryptographic applications of such a function we formulate additional requirements that it has to satisfy. These requirements include resistance to inversion (known as ''one-way property'' or ''preimage resistance''), which makes it unlikely to ''extract'' encoded information out of the quantum state, and resistance to quantum collisions, which means that high similarity of quantum images for different inputs is hardly likely. Below we describe them in more details.

\paragraph{One-way property.}

We present the following definition of a quantum $\delta$-one-way function. Let ${\cal
M}$  be a function ${\cal M} : ({\cal H}^2)^{\otimes s} \to \mathbb X$. Informally speaking ${\cal
M}$ is an ``information extracting'' mechanism that performes some measurement of the state
$\ket{\psi}\in ({\cal H}^2)^{\otimes s}$ and decodes the result into $\mathbb X$.

\begin{definition}\label{one-way-def}
Let $X$ be a random variable 
distributed over $\mathbb X$ $\{ Pr[X=w] : w\in{\mathbb X}\}$. Let $\psi :  \mathbb X \to ({\cal
H}^2)^{\otimes s} $ be a quantum function. Let $Y$ be a random variable  over $\mathbb X$ obtained
by some $\cal M$ that makes some measurement of the encoding $\psi$ of $X$ and decodes the result into
$\mathbb X$. Let $\delta>0$. We call a quantum function $\psi$  a $\delta$-one-way
function  if for any $\cal M$, the probability $Pr[Y  = X]$ that $\cal M$ successfully decodes $Y$
is bounded by $\delta$
\begin{equation}
Pr[Y = X] \le \delta.
\end{equation}

  \end{definition}
For the cryptographic purposes it is natural to assume (and we do this in the rest of the paper)
that the random variable $X$ is uniformly distributed.

A quantum state of $s\ge 1$ qubits can ``carry''  an infinite amount of information. On the other
hand, the fundamental result of quantum information known as the Holevo theorem
\cite{Holevo:1973:bound} states that a quantum measurement can only give $O(s)$ bits of information
about such a state. We use here the following version
\cite{Nayak:1999:Bounds-for-Quantum-Automata} of the Holevo theorem:

\begin{property}\label{preimage-bound}
Let $X$ be a random variable uniformly distributed over the finite set $\mathbb X$. Let $\psi :  \mathbb X \to ({\cal H}^2)^{\otimes s} $ be a quantum function. Let $Y$
be a random variable over $\mathbb X$ obtained by some $\cal M$ that makes some measurement of the
encoding $\psi$ of $X$ and decodes the result into $\mathbb X$. Then the probability of correctly
decoding $Y$ is given by
\begin{equation}
Pr[Y=X] \le \frac{2^s}{\left|\mathbb X\right|}.
\end{equation}
\end{property}

\paragraph{Collision resistance.}

The following definition describes the property of collision resistance for a quantum function:
\begin{definition}
\label{QHF}
Let $\varepsilon >0$. We call a quantum function
 $ \psi : {\mathbb X} \to ({\cal H}^2)^{\otimes s} $
a $\varepsilon$-collision-resistant function if
 for any pair $w,w'$ of different inputs,
\begin{equation}
\left|\braket{\psi(w)}{\psi(w')}\right| \le \varepsilon.  \quad
\end{equation}

\end{definition}

Note that the above inequality means near-orthogonality 
of quantum
states $\ket{\psi(w)}$ and $\ket{\psi(w')}$. It is well-known that orthogonality of quantum states
provides their distinguishability. In the context of quantum functions near-orthogonality
means high collision resistance. That is, let us denote by $Pr_{T}[v=w]$ a probability that
some test ${ T}$ for given quantum states $\ket{\psi(v)}$ and $\ket{\psi(w)}$ outputs the result
``$v=w$''.
For example, the well-known {\em SWAP}-test  \cite{Buhrman:2001:Fingerprinting} gives this result
with the probability
\begin{equation}
Pr_{swap}[v=w] \le \frac{1}{2}(1+ \varepsilon^2),
\end{equation}
while the {\em REVERSE}-test \cite{Ablayev-Vasiliev:2014:Crypto-Q-Hashing}, \cite{GC:2001:Quantum-Digital-Signatures} gives
\begin{equation}
Pr_{reverse}[v=w] \le \varepsilon^2.
\end{equation}

Note also, that this $\varepsilon$-collision-resistance property also corresponds to the classical \emph{Second pre-image resistance}, since one cannot find two different messages for which the SWAP-test or REVERSE-test would
erroneously output equality with probability close to 1.

There is a known lower bound by Buhrman et al. \cite{Buhrman:2001:Fingerprinting} for the size of
the sets of pairwise-distinguishable states: to construct a set of $\left|\mathbb X\right|$ quantum states with
pairwise inner products below $\varepsilon$ we will need at least $\Omega(\log(\log{\left|\mathbb X\right|}/\varepsilon))$ qubits. This implies that an  $\varepsilon$-collision-resistant quantum function requires at least $s=\Omega(\log\log|{\mathbb X}|-\log{\varepsilon}))$ qubits. The similar lower
bound of $\log\log{K} - c(\varepsilon)$ was proved by a different method in
\cite{Ablayev-Ablayev:2015:e-universal-Quantum-Hashing}.
Thus, from the Property \ref{preimage-bound} it is follows that an $\varepsilon$-collision-resistant quantum function can be $\delta$-one-way for $\delta=\Omega\left(\log|{\mathbb X}|/\varepsilon^2|{\mathbb X}|\right)$.

The above two definitions and considerations lead to the following formalization of the quantum
cryptographic (one-way and collision resistant) hash function:
\begin{definition}\label{main_def}
Let $s\ge 1$, $\delta\in(0,1]$ and $\varepsilon\in[0,1)$.  We call a function
 $ \psi : {\mathbb X} \to ({\mathcal H}^2)^{\otimes s}
 $
a quantum $(\delta,\varepsilon)$-hash function  if $\psi$
 is   $\delta$-one-way  and
   $\varepsilon$-collision-resistant function.
\end{definition}

\paragraph{The trade-off between one-way property and collision resistance.}
We have shown earlier \cite{Ablayev-Ablayev-Vasiliev:2016:Balanced-Quantum-Hashing} that one-way property and
collision resistance lead to the contradictory requirements on the size of the quantum hash so that the
``more'' a quantum function is one-way the ``less'' it is collision resistant and vice
versa.

\begin{example}\label{enc1}
We encode a  word  $w\in \{0,1\}^k$ into the one-qubit state:
\[{\ket{\psi(w)}= \cos\left(\frac{\pi w}{2^k}\right)\ket{0} +
  \sin\left(\frac{\pi w}{2^k}\right)\ket{1}}.  \]
 Here we treat $w=w_{k-1}\dots w_0$ also as a number
    $w=w_0+w_12^1+\dots +w_{k-1}2^{k-1}$.
This function has good one-way property with $\delta=\frac{2}{2^k}$, but has poor collision resistance with $\varepsilon=\cos\left(\pi/ {2^{k}}\right)$.
\end{example}

\begin{example}\label{enc2}
We encode a  word  $w\in \{0,1\}^k$ into the $k$-qubit state:
\[{\ket{\psi(w)}= \ket{w}}.  \]
This function has $\delta=1$ (no preimage resistance) and collision resistance with
$\varepsilon=0$ (perfect resistance).
\end{example}

\section{Preparing single-photon states with orbital angular momentum}\label{sec:Experiment}

For the quantum hash to be implemented, we take advantage of conditional preparation of single-photon states via spontaneous parametric down-conversion (SPDC) \cite{QE1977,PhysRevLett1986}. In the process of SPDC, a strong pump field is directed into a quadratic nonlinear medium, where the pump photons are anihilated with the creation of pairs of photons, generally called signal and idler, which satisfy the conditions of phase matching $\omega_p=\omega_i+\omega_s$ and  $\vec{k}_p=\vec{k}_i+\vec{k}_s$. Here $\omega_n$ and $\vec{k}_n$ are the frequency and wave vector of the signal photon ($n=s$), idler photon ($n=i$) and pump photon ($n=p$). Detecting one of the emitted photons (for example, signal) clearly indicates the presence of a second one in another channel (idler) so that we obtain a heralded single-photon source (see \cite{Castelletto:2008jj,Eisaman:2011cc} for review).

In the case of collinear SPDC, when all the interacting fields propagate in the same direction, an additional conservation low for OAM of the fields takes place \cite{mair2001entanglement,franke2002two,PhysRevA_69_023811,Osorio:2008gh,IbarraBorja:19} that can be written as
\begin{equation}\label{eq:ell}
\ell_p=\ell_i+\ell_s,
\end{equation}
where $\ell _n$ is the topological charge of the OAM for $n$th photon ($n=\{p,s,i\}$), which corresponds to a vortex wavefront described by the azimuthal phase factor $\exp(i\ell_n\varphi)$. 
The OAM conservation, Eq.~(\ref{eq:ell}), allows us to control the spatial state of the heralded photon by modulating the pump field provided that heralding photon is detected in a fixed spatial state. In particular, if the signal photons are detected with $\ell_s=0$, then $\ell_i=\ell_p$ and the spatial structure of the idler photon reproduces that of the pump field. 
In doing so, we can conditionally prepare single-photon qubits encoded via OAM with a maximum heralding efficiency since heralded photons need no additional spatial transformation. In the present work, we realize this approach by preparing and measuring OAM carrying beams in the basis of Laguerre--Gaussian (LG) modes (see Appendix A).

\begin{figure}[t]
\includegraphics[scale=0.43]{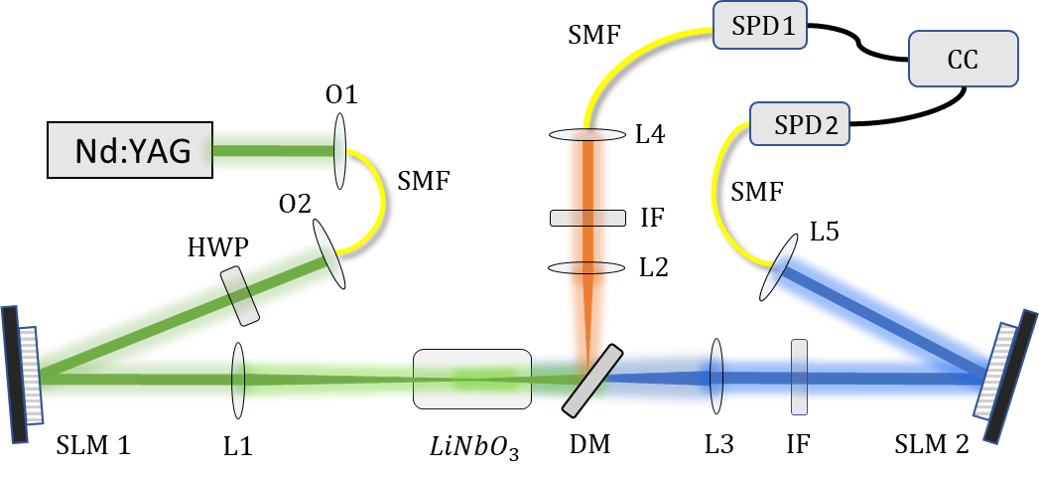}
\caption{\label{fig:epsart} Schematic of the experimental setup (see text for details).}
\end{figure}

The experimental setup is schematically depicted in Fig.~\ref{fig:epsart}. We used a  10-mm-long periodically-poled lithium niobate crystal doped with magnesium oxide ($\text{LiNbO}_3\text{:MgO} (5$\%$)$) for a collinear frequency non-degenerate type-0 phase matching as a source of photon pairs.
As a pump, we use the second harmonic of a cw neodymium laser operating at a wavelength of 532 nm, which results in generating photons at the central wavelength of 860 nm (signal) and 1480 nm (idler). The pump field was spatially filtered by a single-mode fiber (SMF) interfaced by two objectives (O1 and O2). To prepare required spatial states of the pump field we take advantage of a phase holography technique developed in \cite{bolduc2013exact}. For the pump field to be transferred into an LG mode, a spatial light modulator (SLM~1, Holoeye) was used. 
In doing so, a half-wave plate (HWP) was used to optimize the field polarization with respect to the SLM~1.  A required structured pump field is obtained in the first diffraction order and is focused into a crystal using a lens (L1) with a focal length of 250 mm. The signal and idler photons, generated at the wavelength of 860 nm and 1480 nm, respectively, are separated by a dichroic mirror (DM).
In the signal arm, the photons are collimated using a lens L2 with a focal length of 150 mm and then are coupled with an aspheric lens (L4) with a focal length of 11 mm into a single-mode fiber (SMF). As a result, the detected photons have only zero OAM ($\ell_s$=0). In the idler arm, the photons collimated using a lens L3 with a focal length of 100 mm fall on a second spatial light modulator (SLM 2, Holoeye) and have been transformed are coupled with aspheric lens (L5) into a single-mode fiber (SMF). Interference filters IF in both channels were used to filter out undesirable radiation including the pump field. Such a system allows us to measure the spatial states of the idler photons using the phase-flattening technique \cite{mair2001entanglement}.

The focal lengths of L1--L3 were chosen to optimize coupling and approach single-Schmidt mode regime of SPDC \cite{Kovlakov_2017}. To this end, the length of the crystal $L$ should be twice the the Rayleigh range of the Gaussian pump beam $z_R=\pi w_p^2/\lambda_p$, where $\lambda_p$ denotes the wavelength of the pump field, so that the required pump beam waist is estimated to be 
$w_p = \sqrt{L / k_p} \approx  29\,\mu\rm{m}$. For the optimal detection of the down-converted modes, their beam waists should be $w_{s,i} = \sqrt2 w_p \approx 41\,\mu\rm{m}$. In the present experiment, we had $w_p = 33\,\mu\rm{m}$, $w_s = 41\,\mu\rm{m}$ and $w_i = 47\,\mu\rm{m}$.

Detection of photons was carried out using photodetectors SPD1 (SPCM AQR-14, PerkinElmer) operating in the free-running mode with an efficiency of $45\%$, dark count rate of the order of 2 kHz and dead time of 150 ns, and SPD2 (ID Quantique 210) operating in the external gating mode with an efficiency of $25\%$, dark count rate of the order of 10 kHz and dead time of 14 $\mu$s. The signals from both detectors analyzed by the coincidence counting curcuit (CC, ID Quantique 800). The photon-pair generation rate, with allowance for losses in the optical system, was found to be 176 kHz/mW. The heralding efficiency, which is defined as the ratio of the count rate in the idler (heralded) channel to the count rate in the signal (heralding) channel, turned out to be 11\%. 
 
\begin{figure}[h]
\includegraphics[scale=0.4]{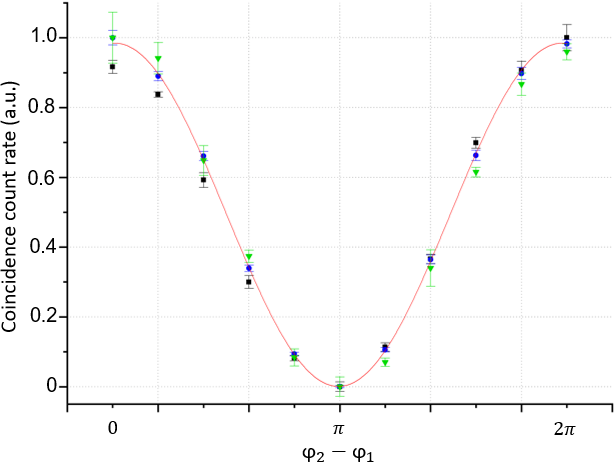}
\caption{\label{fig:sine} Coincidence count rate as a function of the phase difference $\varphi_2-\varphi_1$, where $\varphi_1$ corresponds to the qubit state prepared by SLM 1 and $\varphi_2$ corresponds to the qubit state for which phase flattering is achieved by SLM 2. Solid line is the theoretical dependence; black square, blue circles, green triangles correspond to $l=1,2$ and 3, respectively.}
\end{figure}

In this work, we prepare a single-photon qubits as a superposition of two $\text{LG}_0^\ell$ modes with opposite values of the topological charge $\ell$ in order to cancel contributions of the Gouy phase. In particular, we focus on the equally weighted superpositions that have the following form:
\begin{equation}\label{eq:OAM-states)}
\ket{\psi}=\frac{1}{\sqrt{2}}(\ket{\ell}+e^{i\varphi}\ket{-\ell}),
\end{equation}
where $\ket{\ell}$ denotes a single-photon state corresponding to $\text{LG}_0^\ell$ mode, and $\varphi$ is a relative phase. To illustrate preparing and measuring such qubit states for different values of $\ell$, we generate a simple superposition $\ket{\psi_1}$ with $\varphi_1=0$ using SLM 1 and project the state of heralded photons onto a state $\ket{\psi_2}$ with $\varphi_2\in [0,2\pi]$ using SLM 2, which provides phase flattering for the corresponding superposition of LG modes. Fig.~\ref{fig:sine} shows the resulting coincidence count rate depending on the phase difference, which should be described by $\cos^2[(\varphi_2-\varphi_1)/2]$. We clearly see that high visibility is achieved for different values of $\ell$ so that a close to zero coincidence count rate is observed for projecting on the orthogonal qubit state (when $\varphi_2-\varphi_1=\pi$) independently of the topological charge value, and no phase shift due to the Gouy effect is observed.
The symmetric intensity distribution of the generated spatial modes is extremely sensitive to positioning of the incident beam on the hologram. In our experiment, the positioning accuracy was about $8\,\mu\rm{m}$. The single-photon state preparation rate is limited by SLM switching rate which is equal to 60 Hz in the present experiment.

\section{OAM-based Quantum Hashing}

We propose a quantum hash function that exploits superposition states of OAM (\ref{eq:OAM-states)}). 
That is, the classical input $x\in \{0, 1,\ldots, q-1\}$ is encoded by the relative phase of $s$ single-photon states:
\begin{align}\label{eq:OAM-Hashing)}
& \ket{\psi_j(x)}=\frac{1}{\sqrt{2}} (\ket{\ell}+e^{i\frac{2\pi b_jx}{q}}\ket{-\ell}),\\
& \ket{\psi(x)}=\ket{\psi_1(x)}\otimes\cdots\otimes
\ket{\psi_s(x)},
\end{align}
where $b_j\in \{0, 1,\ldots, s\}$ are numeric parameters of the quantum hash function that provide its
collision resistance.

The main idea behind quantum hashing is to provide the minimal fidelity of different quantum hash
codes (collision resistance) with the minimal possible number of qubits (that affects the one-way property). As mentioned earlier there is always a trade-off between these two properties so that to balance them in a quantum hash function is an important task.

The fidelity of $\ket{\psi(x_1)}$ and $\ket{\psi(x_2)}$ is given by 
\begin{align}\label{eq:Hashing-Overlap)}
\left|\braket{\psi(x_1)}{\psi(x_2)}\right|^2&=\frac{1}{{2}^{2s}}\prod\limits_{j=1}^s\left|1+e^{i\frac{2\pi b_j(x_2-x_1)}{q}}\right|^2\\
&=\frac{1}{{2}^{s}}\prod\limits_{j=1}^s\left(1+\cos{\frac{2\pi b_j(x_2-x_1)}{q}}\right)
.
\end{align}

Thus, the set of parameters $B=\{b_1,\ldots, b_s\}$ should give the minimal fidelity of all pairs of unequal $\ket{\psi(x_1)}$ and $\ket{\psi(x_2)}$, i.e.
\begin{equation}
B=\underset{b_1,\ldots,b_s}{\operatorname{argmin}} \max\limits_{x\neq 0}\frac{1}{{2}^{s}}\prod\limits_{j=1}^s\left(1+\cos{\frac{2\pi b_jx}{q}}\right).
\end{equation}

The search for this set may be done with various heuristic techniques (genetic algorithms, simulated annealing, etc.) or even with exhaustive search for small values of $s$.


In the present experiments we test two basic procedures that may be incorporated into various computational and cryptographic protocols. 
These are generating a quantum hash and its verification that checks whether two hashes are equal or not.

The generation of the quantum hash $\ket{\psi(x)}$ for an input $x$ is given by the experimental technique described in Sec.~\ref{sec:Experiment} and is just preparing a sequence of single-photon qubits with the relative phase $\varphi_j$ for the $j$th qubit equal to $2\pi b_jx/{q}$. 

The ideal quantum experiment that verifies a quantum hash can be set as following:
\begin{itemize}
    \item We receive a quantum hash of some generally unknown value $x_1$ as a sequence of $s$ single photons in the overall state
    $$\ket{\psi(x_1)}=\ket{\psi_1(x_1)}\otimes\cdots\otimes \ket{\psi_s(x_1)},$$
    and we want to check whether $x_1$ equals some predefined $x_2$ or not;
    \item The $j$th qubit is expected to be in the state
    $$\ket{\psi_j(x_1)}=\frac{1}{\sqrt{2}} (\ket{\ell}+e^{i\frac{2\pi b_jx_1}{q}}\ket{-\ell}),$$
    and we perform a measurement that projects $\ket{\psi_j(x_1)}$ to either $\ket{\psi_j(x_2)}$ or 
    $$\ket{\psi'_j(x_2)}=\frac{1}{\sqrt{2}} (\ket{\ell}+e^{i\frac{2\pi b_jx_2}{q}+\pi}\ket{-\ell}),$$
    which is orthogonal to $\ket{\psi_j(x_2)}$;
    \item The measurement is performed by a state selective beam splitter and two single-photon detectors that click on $\ket{\psi_j(x_2)}$ or $\ket{\psi'_j(x_2)}$, which corresponds to the outcome ``$\ket{\psi_j(x_1)}=\ket{\psi_j(x_2)}$'' or ``$\ket{\psi_j(x_1)}\neq\ket{\psi_j(x_2)}$'', respectively;
    \item If $x_1=x_2$, the detector of the output $\ket{\psi_j(x_2)}$ would always click, while the other detector would never click;
    \item If $x_1\neq x_2$, each of the detectors might click, but the probability of erroneous outcome ``$x_1=x_2$''  is bounded by the construction of the quantum hash function $\psi$;
    \item If none of the detectors had clicked, then the qubit is lost, and we either request its resending or accept the higher error probability;
    \item If all of $s$ measurements end up with the outcome ``$\ket{\psi_j(x_1)}=\ket{\psi_j(x_2)}$'', then the final result of experiment is considered to be ``$x_1=x_2$''. Otherwise, if at least one qubit leads to ``$\ket{\psi_j(x_1)}\neq\ket{\psi_j(x_2)}$'', then the overall result is also ``$x_1\neq x_2$''.
\end{itemize}

 Obviously, when $x_1=x_2$ the states are the same and the test should always confirm this. However, for
$x_1\neq x_2$ both test outcomes are possible. The probability of erroneously confirming the equality of two different hashes is bounded by the fidelity of corresponding states. Since 
\begin{equation}
    \left|\braket{\psi(x_1)}{\psi(x_2)}\right|^2=\left|\braket{\psi(0)}{\psi(x_2-x_1)}\right|^2,
\end{equation}
it is sufficient to consider the case of comparing $\ket{\psi(x)}$ to $\ket{\psi(0)}$ for $x\neq0$.

The state selective beam splitter mentioned above can be realized by generating appropriate holograms on SLM 2 as shown in \cite{li2015simultaneous}. Since it is difficult to
prepare arbitrary long hash functions and measure the error rate
within them, in order to characterize the protocol statistically, we
repeat the measurements many times, i.e.
 to distinguish the outcomes ``$x_1=x_2$'' and ``$x_1\neq x_2$'' we measure the coincidence count rate between the signal and idler channels. The first corresponds to a heralding photon while the second to a photon of the hash function. A detector click in the first channel means that the second photon has been prepared in the state $\ket{\psi(x_1)}$ and a detector click in the second channel proves that the second photon prepared in such a state has been successfully projected on the state  $\ket{\psi(x_2)}$.  Therefore, the coincidence count rate is proportional to the probability $\left|\braket{\psi(x_1)}{\psi(x_2)}\right|^2$, which we are interested in. In doing so, we compare different
quantum hashes in the worst-case scenario, i.e. we perform the test for
$\ket{\psi(x_{max})}$ and $\ket{\psi(0)}$, where
\begin{equation}
x_{max}=\underset{x\neq 0}{\operatorname{argmax}} \frac{1}{{2}^{s}}\prod\limits_{j=1}^s\left(1+\cos{\frac{2\pi b_jx}{q}}\right)
\end{equation}
for a given optimal (quasi-optimal) set $B=\{b_1,\ldots, b_s\}$.

The protocol starts with an initializing step on which we adjust the coincidence count rate for the ``yes''-answer (``$x_1=x_2$'').  We perform about 100 measurements of equal states $\ket{\psi(0)}$ and $\ket{\psi(0)}$ and calculate the average coincidence count rate. We iterate this step about 10 times and pick the minimal value as the borderline between ``yes'' and ``no''. This corresponds to the worst-case scenario since it would be easier to confuse equal and unequal states.

Then we perform the protocol for $q=512$ and different hash sizes $s$ between 2 and 8. The required level of accuracy in preparing and measuring superpositions given by Eq.~(\ref{eq:OAM-Hashing)}) has been checked by quantum tomography as described in Appendix B. For $s$ qubits and $\left|\mathbb X\right|=q$ we obtain the inversion probability bounded by $\delta=2^s/q$. 
For each qubit we calculate the average coincidence count rate when comparing to the state $\ket{\psi(0)}$. Thus, the ratio between this quantity and the lower bound from initializing step gives us the statistical estimate of the fidelity between this state and $\ket{\psi(0)}$, which corresponds to the probability of erroneously accepting unequal states.

\begin{figure}[t!]
\includegraphics[scale=0.33]{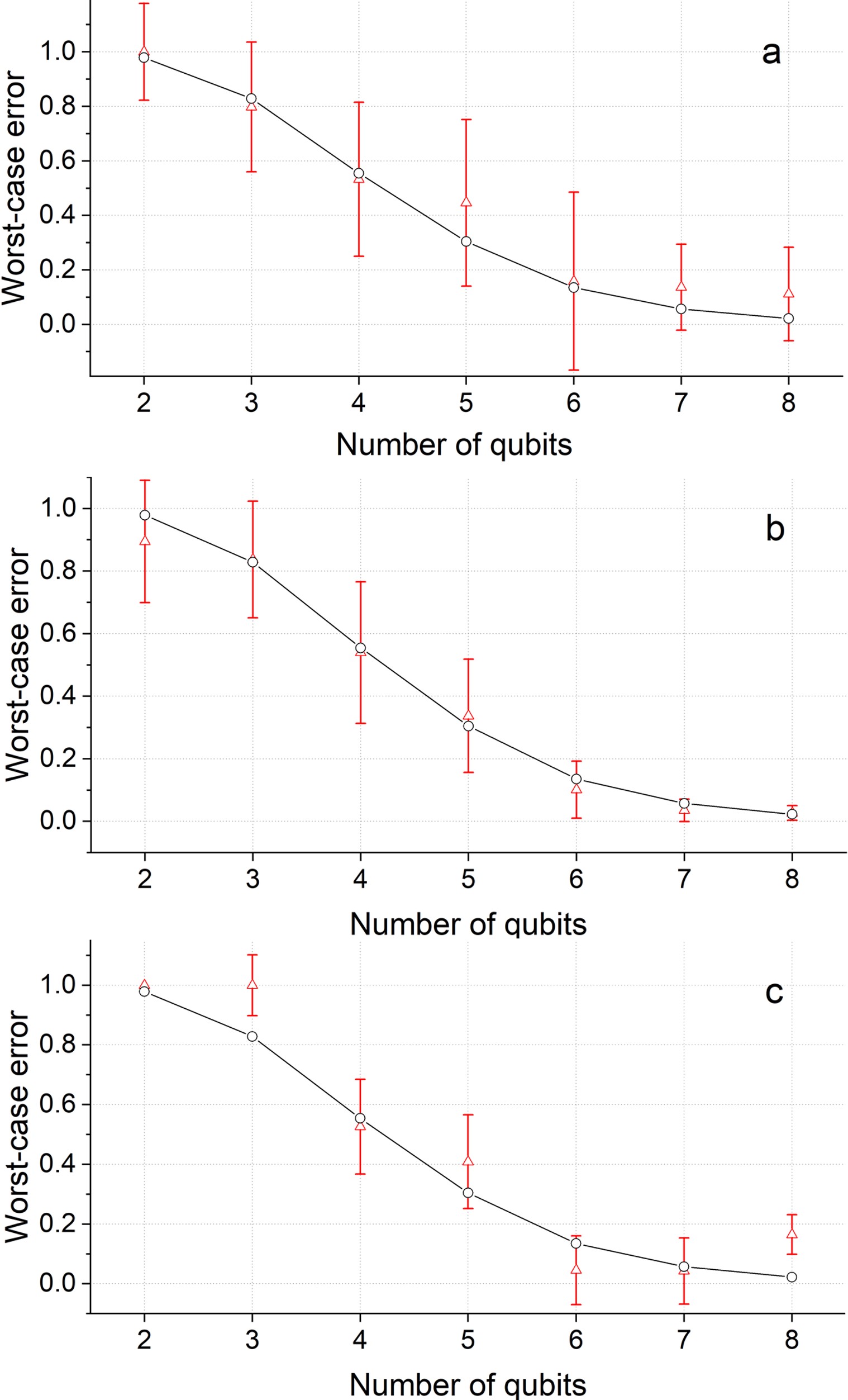}
\caption{\label{fig:Verification-Test}
Comparing experimental and theoretical error rates for the worst-case scenario and different OAM basis states. The open circles (triangles) show theoretical (experimental) values for the cases $l=1$ (a), $l=2$ (b) and $l=3$ (c).}
\end{figure}

Fig.~\ref{fig:Verification-Test} shows the comparison of experimental and theoretical error rates for the worst-case scenario and different OAM basis states. The corresponding data are present in Appendix C. It can be seen that experimental values agree with the theoretical prediction independently of the topological charge value. Therefore, multiplexing different OAM bases is possible making quantum hash function preparing and measuring faster. 

The experimental results suggest that the considered technique can be useful even for small sizes of input and output states. For instance, for $s=5$ we ``compress'' the 9-bit input strings into 5-qubit states, which provides some kind of the balance between inversion probability bounded by $32/512=1/16$ and collision probability bounded by $0.31$. The former suggests the usage in cryptographic scenarios, while the latter is sufficient for computations with one-sided errors (e.g., deciding languages from one of the central complexity classes of practical problems coRP \cite{Wegener:2005:Complexity-Theory} requires the one-sided error bounded by $1/2$).

\section{Conclusion}
In this paper, we have demonstrated a proof-of-principle implementation of quantum hashing based on single-photon superposition states with OAM.
An important result is the experimental verification of the collision resistance of quantum hashing (i.e., the coincidence of different hashes during their measurement) for various number of qubits (which affects the one-way property). We took the lowest average coincidence rate as the borderline between ``yes'' and ``no'', which corresponds to the worst-case scenario. Experimental results have confirmed theoretical estimates of the error probability dependence on the number of qubits in a quantum hash.

Overall, experimental results for different bases of OAM states confirm theoretical estimates and show the potential performance of this approach in computational and cryptographic scenarios. The possibility of multiplexing different bases of OAM states can speed up the communication by spreading quantum hash states between them, i.e. disjoint subsets of single-photon states can be transferred in parallel by these technique. Alternatively, it can provide redundancy for more tolerance to errors and qubit losses.

\begin{acknowledgments}
The experimental part of the work was supported by the Russian Science Foundation (project No. 19-19-00656). The numerical analysis was made within the government assignment for FRC Kazan Scientific Center of RAS.
\end{acknowledgments}

\section*{APPENDIX A}
The amplitude distribution of Laguerre--Gaussian (LG) modes in cylindrical coordinates is given by \cite{Barnett:2017bq}
\begin{eqnarray}
  \label{eq:Lg}
    \text{LG}_p^\ell(\rho,\phi,z) =
    &&\sqrt{\frac{2p!}{\pi(|\ell|+p)!}} \frac{1}{w(z)} 
    \left[\frac{\sqrt{2}\rho}{w(z)}\right]^{|\ell|}L_p^{|\ell|}\left[\frac{2\rho^2}{w^2(z)}\right] \nonumber\\
      &&\exp\left[-\frac{\rho^2}{w^2(z)}\right]  \exp\left[ -i k \frac{\rho^2z}{2(z_R^2+z^2)}\right] 
   \nonumber\\
      && \exp(i\ell\varphi) \exp [-i\Psi(z)],
\end{eqnarray}
where $\ell$ is the azimuthal index giving an OAM of $l\hbar$ per photon, which is also called topological charge or the OAM, $p$ is the radial index (the number of radial nodes in the intesity distribution), $L^{|\ell|}_{p}$ is an associated Laguerre polynomial, $w(z)=w(0)[(z^2+z_R^2)/z_R^2]^{1/2}$ is the radius of the beam, $z_R=k w(0)^2/2$ is the Rayleigh range, $\Psi(z)=[(|\ell|+2p+1) \arctan({z}/{z_R})]$ is the Gouy phase.

\section*{APPENDIX B}

Quantum hashes are generated by preparing a sequence of $s$ single-photon qubit states given by Eq.~(\ref{eq:OAM-Hashing)}) with the phase $\varphi_j$ for the $j$th qubit state ($j=1,...,s$) equal to $ 2 \pi b_j x / q $, where $b_j,x\in \{0, 1,\ldots, q-1\}$ and  $q=2^s$. However, due to imperfections of the experimental setup, one would expect a limited accuracy of the phase control. To check the latter, we perform quantum tomographic measurements for the case $s=8$ and $q=512$, where the minimal value of the phase difference between qubit states is $\pi/256$. In doing so, we take advantage of the  quantum tomography approach developed in \cite{agnew2011tomography}.

\begin{widetext}

As an example, we carry out quantum tomographic measurements for a qubit state $\ket{\psi}=\frac{1}{\sqrt{2}}(\ket{\ell}+e^{i\varphi}\ket{-\ell})$ with $\ell = 2$ and $\varphi = 2\pi/3$ (or $120^\circ$). 
Such a qubit state is described by the density matrix 
\begin{align}
  \label{eq:2pi/3theor}
\rho_{\varphi}^{\rm{theor}}=
&\left(\begin{array}{cccc}
0.500 & -0.250 + 0.433i \\
-0.250 - 0.433i  &  0.500\\
\end{array}\right).
\end{align}
The reconstructed state proves to be 
\begin{align}
  \label{eq:2pi/3}
\rho_{\varphi}^{\rm{exp}}=
&\left(\begin{array}{cccc}
0.500 \pm 0.004 & -0.251 \pm 0.001 + (0.432 \pm 0.001)i \\
-0.251 \pm 0.001 - (0.432 \pm 0.001)i  &  0.499 \pm 0.004\\
\end{array}\right),
\end{align}
which corresponds to the observed qubit phase $\varphi_{\rm{exp}} = (120.14 \pm 0.15)^\circ$.

Next, we shift the qubit phase by a minimum step equal to $\delta\varphi=\pi/256$ (or $0.703^\circ$) and measure the new density matrix that is expected to be
\begin{align}
 \label{eq:pi/256theor}
\rho_{\varphi-\delta\varphi}^{\rm{theor}}=
&\left(\begin{array}{cccc}
0.500 & -0.245 + 0.436i \\
-0.245 - 0.436i  &  0.500\\
\end{array}\right).
\end{align}
The result is 
\begin{align}
 \label{eq:pi/256}
\rho_{\varphi-\delta\varphi}^{\rm{exp}}=
&\left(\begin{array}{cccc}
0.499 \pm 0.002 & -0.241 \pm 0.001 + (0.432 \pm 0.004)i \\
-0.241 \pm 0.001 - (0.432 \pm 0.004)i  &  0.500 \pm 0.002\\
\end{array}\right),
\end{align}
which corresponds to the observed phase $(\varphi-\delta\varphi)_{\rm{exp}} = (119.21 \pm 0.2)^\circ$.

Based on the data obtained, we observe the relative phase shift $\delta\varphi_{\rm{exp}}= (0.93 \pm 0.25)^\circ$, which allows us to conclude that the qubit states are controlled with the sufficiently high accuracy in phase.

\end{widetext}

\section*{APPENDIX C}

The experiment was set for the values of $x=x_1-x_2$ that correspond to the worst pairs of $x_1, x_2$, i.e., those with maximal fidelity between the states $\ket{\psi(x_1)}$ and $\ket{\psi(x_2)}$. This is equivalent to comparing the value of $x=x_1-x_2$ to 0, since this is also the worst case.

Table \ref{tab:collision-resistance-bounds} shows the worst-case values of $x$ that have been calculated for various sizes of a quantum hash and the corresponding values of theoretical and experimental bounds on collision resistance.

\begin{table*}
\caption{\label{tab:collision-resistance-bounds}The worst-case values of $x$ for various sizes of a quantum hash and the corresponding values of theoretical and experimental bounds on collision resistance.}
\begin{ruledtabular}
\begin{tabular}{cccccc}
Hash size  &
Worst-case  &
Theoretical  &
  &
Experimental results  &
 
          \\
(number of qubits) &
value of $x$ &
error bound &
$l=1$ &
$l=2$ &
$l=3$\\
\hline

2 &
18 &
0.9784 &
1 &
0.89476 &
1 \\
 
3 &
163 &
0.8286 &
0.797747 &
0.837321 &
1 \\

4 &
4 &
0.5544 &
0.532488 &
0.539566 &
0.526449 \\

5 &
207 &
0.3047 &
0.44643 &
0.337337 &
0.40883 \\

6  &
71 &
0.1356 &
0.15882 &
0.101393 &
0.045501 \\

7 &
39 &
0.0574 &
0.136674 &
0.035423 &
0.042659 \\

8 &
28 &
0.0222 &
0.11205 &
0.026552 &
0.165335\\
 
\end{tabular}
\end{ruledtabular}
\end{table*}


\bibliography{apssamp,references}

\end{document}